\documentclass{ws-acs}
\usepackage{graphicx}
\usepackage{amsmath}
\usepackage{amssymb}
\usepackage{color}
\def\slaninafigdir{.}
\begin{document}
\title{%
EIGENVECTOR LOCALIZATION AS A TOOL TO STUDY SMALL COMMUNITIES IN ONLINE SOCIAL NETWORKS
}
\author{%
FRANTI\v{S}EK SLANINA
}
\address{%
Institute of Physics,\\
 Academy of Sciences of the Czech Republic,\\
 Na~Slovance~2, CZ-18221~Praha,
Czech Republic
\\
slanina@fzu.cz
}
\author{%
ZDEN\v{E}K KONOP\'ASEK
}
\address{%
Center for theoretical study,\\
 Charles University in Prague / Academy of Sciences of the Czech Republic,\\
 Jilsk\'a 1, Praha, Czech Republic
\\
zdenek@konopasek.net
}

\maketitle
\date{\today}%
\begin{abstract}
We present and discuss a mathematical procedure for identification of
small ``communities'' or segments  within large bipartite
networks. The procedure
is based on spectral analysis of the matrix encoding network
structure. The principal tool here is localization of  eigenvectors of
the matrix, by means of which the relevant network segments become visible. 
We exemplified our approach by analyzing 
the data related to product reviewing on Amazon.com.
We found several  segments, a kind of hybrid communities of densely
interlinked reviewers and products, which we were able to meaningfully
interpret in terms of the type and thematic categorization of reviewed
items.
The method provides a complementary approach 
to other ways of community detection, typically aiming at
identification of large network modules.
\end{abstract}

\keywords{%
{social network}%
;
{random matrix}%
;
{internet}%
}%

\section{Introduction}
The complexity of our societies is studied by social analysts in
various ways. Qualitative inquiries and case studies usually put
little emphasis on formalized description, partly to avoid
oversimplification, or even trivialization of the phenomena under
study. On the other side, sophisticated mathematical procedures are
increasingly used in order to grasp complexity in a specific way, as a
formalized property of larger systems. One of the branches of the
latter stream is represented by the  analysis of social networks using
mathematical theory of graphs. Our approach  adheres precisely to this
field of research and yet, it follows slightly different direction
than most efforts in contemporary network analysis.  

The purpose of this paper is twofold. First, we want to present a
specific solution to a rather standard problem of social network
analysis, which is identification of communities within complex
networks. Second, we want to discuss some alternative perspectives on
the concept of ``social network''. We suggest that our method might
provide a suitable tool for empirical research in respective
directions, enabling the analyst to determine those ``hot spots'' within
the network that usually escape attention.

To make the wider methodological context of our paper clearer, let us
start with some notes on networks and network analysis in contemporary
sociology. 
The use of the term  ``network''  in contemporary
sociology vary from loose 
metaphors \cite{castells_2000} to rather specific and technical
meanings \cite{degenne_forse_1994,scott_2000}, compatible with the
network science as understood in mathematics or physics. 

Social network analysis has a complex history, with roots involving
the sociometric analysis of Moreno in the beginning of 20th century,
the Harvard researchers of the 1930s and 1940s who studied
interpersonal configurations and cliques and, finally, the group of
anthropologists based in Manchester who, roughly in the same time,
instead of emphasizing
integration and cohesion as their predecessors, focused on conflict
and change, see \cite{scott_2000}, pp. 7-37. In 1960s, a key turn to
mathematization occurred, which gave this field a new impulse and high
ambitions. 
Today, encouraged by the rise of interest in networks in
other scientific disciplines, social network analysis is sometimes
seen as an approach that may entirely redefine the social sciences,
while integrating them into a broader interdisciplinary research
program \cite{christakis_fowler_2009}. 
Formalized analytical procedures hugely contributed to the fact that
social network analysis has 
become firm basis for social science discussions
\cite{wasserman_faust_1994}. 
However, integration of
mathematical analytic thinking with sociological imagination is 
{ an
intricate} task. As noted by
\cite{emirbayer_goodwin_1994}, the application of  
formalized methods of social network analysis is often marked by
neglecting substantive and theoretical sociological
consequences. Also, despite the growing popularity of mathematical modeling,
qualitative, {or ethnographic} studies of
{``network sociality'' \cite{wittel_2001}} keep their
relevance, hand in 
hand with quantitative approaches.

Given this complicated background, our aim, in this paper, is rather
modest. We want to introduce
and illustrate a new mathematical method for 
identification of small parts of complex networks with higher level of
commonalities and for studying their basic formal properties. As an
example and possible field of application 
we have chosen networks of product reviewing on the Amazon.com
portal.  Here, the simplest possible ties structuring the network are
the connections established by two reviews written on the same
product. In other words, what reviewers may have in common is the
product reviewed by them. The configurations when one product, e.g., a
book or a CD, is reviewed by two or more reviewers are frequent, of
course, and not much special. But if the same reviewers are similarly
connected via some other items too, the situation gets more
exciting. We can assume that network segments with higher density
of such links represent small communities of reviewers with similar
interests. {Our first and main objective is to find these small communities.}

Identification of such small-size groupings has always been one of the
key tasks in social network analysis.  Identification of these network segments
is an interesting empirical finding in itself. Other times, however,
the need to focus on smaller network segments is rather methodological
than substantial or theoretical: for instance, when David and Pinch
\cite{david_pinch_2008}  analyzed the phenomenon of review plagiarism on Amazon.com,
they had to ``localize'' the phenomena in order to make it better
graspable in detail. Thus, they had to reduce their sample while
focusing on those segments of the vast amount of data available in
which reviewed products were
 ``somewhat similar to one another'' and thus vulnerable to
``recycling'' practices they were interested in. This is a
characteristic situation: complex networks, including the social ones,
are quite often huge, only hardly analyzable in details, with respect
to local deviations or little extremities. This is especially true for
on-line networks. When studying internet-related network structures,
analysts can quickly become overloaded with data and it is difficult
to tell what exactly to look at. The urgent question becomes: how to
locate tiny islands of relevance in the ocean of data archived on the
Internet? We offer a possible mathematical method for precisely such a
task -- a more flexible and background-sensitive one (a ``softer'' one,
in a way) than those already described and used in the field.

We should also stress at this point that our task differs from the
well-studied problem of splitting the network into several modules,
which may perhaps overlap, but as an ensemble, they cover the network
entirely. This is the case  in  metabolic networks, to
mention just one example \cite{pal_der_far_vic_05,gui_ama_05}. 
In our case, we want to focus on
a few ``hot spots'', {small
 communities of interest within the network,} leaving all the rest
behind.

\section{Reviewing networks on Amazon.com as a sociological problem}

Before demonstrating the mathematical procedures, let us also briefly
mention some sociological contexts of the chosen example. Sociologists
have pointed out the increasing importance of the symbolic content
of contemporary economics, which is often associated, among others,
with users' or consumers' active involvement in the complex processes
of product evaluation, qualification, and formation 
\cite{allen_2002,lash_urry_1994}. The role of consumers is particularly enhanced by the 
Internet and by the ways computer technologies shape social
networks \cite{callon_etal_2005}.

A specific and significant part of these processes has been recognized
as ``peer-production of relevance/accreditation'' \cite{benkler_2006}, p. 75 or
simply as ``reputational economy'' \cite{david_pinch_2008}. By reviewing or
commenting items in on-line shops, classifying and rating them,
individual consumers become co-producers of coordinates for others'
economic decision making. They engage in a complex action that cannot
be simply grasped in purely economic terms. 
{As noted by \cite{licoppe_2008}, p. 322, spaces of E-commerce
  are characteristic by countless devices creating  diversity of forms
  of encounter between products and consumers \cite{woolgar_2004}. 
}

User reviews and
comments, for instance, not only serve the purposes of the
seller, but also the consumer community, while simultaneously
being the means for identity building of reviewers themselves
 \cite{foot_2009}. In-depth study
of {all these} complex phenomena seems crucial for
better understanding of 
contemporary ``technological economies'' \cite{barry_slater_2005}.

What kind of groupings are we interested in when we try to locate
segments of reviewers connected by shared reviewed products? We might
be tempted to talk about virtual communities. But these would not be
``virtual communities'' in the usual sense
\cite{rheingold_1993,wellman_etal_1996}; and they would not be ``online
social networks'' as typically  
imagined by social scientists. Both these concepts characteristically refer to groups of
people who directly communicate to each other with the help of
computer networks -- i. e., who know (about) each other and interact by means
of on-line forums, instant messaging, or facebook. Our groups of
Amazon.com reviewers represent a slightly different kind of
entities, though. These people usually do not communicate by addressing each
other and quite often they even do not know each other. They do not
belong to the group by virtue of intentional interaction with the
others, but ``merely'' by doing similar things in a relatively
uncoordinated way: writing reviews on specific products. If
\cite{granovetter_73,granovetter_1983} drew our
attention to the importance of ``weak 
ties'' in social networks, i.e., to the significance of ordinary
informal acquaintances (in comparison to family ties and formal
hierarchies), we could speak here of a kind of ``ultra-weak
ties''. These ties are ``virtual'' in the sense that they are not
``real enough''
in the usual sociological meaning; yet, they are materialized and
 articulated - although not by the reviewers themselves
only. We can clearly see the connections on the Amazon.com web pages: the
reviews of these people are listed together, accompanying the
respective item in the catalog. Moreover, the reviewers do not
become members of this community completely unintentionally, but by
means of quite intentional and personal act of assessing the product
and writing the review. They create the community by highly mediated
interactions, as if ``by the way'', together and via the technology of
on-line shopping.

In the following section we present mathematical tools for
identification, representation and elementary description of precisely
such communities. The proposed procedures may have a value
especially
 in relation to  subsequent sociological analysis of these local anomalies, as its
 precondition.

\section{Finding small communities in networks}

\subsection{Motivation}

The problem of identification parts of the network bearing some
relevant structural information, can be relatively easily formulated
in mathematical terms. 
%
%
%
%
The
methodological problem is, which one of the variety of possible
mathematical formulations of community detection is
suitable for given purpose.
Let us  stress  that we neither aim at improving
the existing schemes nor present an algorithm which should compete
with the established ones. Instead, we are bringing an
alternative scheme which reveals  structures, not
covered by other schemes of community detection. That is why we not
only present a description of the method and its application to
one real-world example, but also spend time putting it into a
wider context of sociological thinking.

\subsection{Background}

For a long time, the standard way of mathematical modeling of social
networks \cite{wasserman_faust_1994,degenne_forse_1994} was the 
``classical'' theory of
random graphs \cite{bollobas_85,diestel_00} initiated by the
work by  Erd\H os and R\'enyi \cite{erd_ren_59}.
However, in the last decade a new class of networks became studied
 and the name ``complex networks'' became common
  denomination for them 
\cite{bar_alb_99,alb_bar_01,strogatz_01,dor_men_01,bor_sch_03,dor_men_03,new_bar_wat_06}. 
Compared with the ``classical'' models of
  random networks, they grasp much better the networks found in
  reality and at the same time their models are much more involved
  than bare random dropping of edges as in the Erd\H os-R\'enyi 
  model. The most immediate characteristics common to the complex
  networks is their  degree distributions with power-law tails
  \cite{bar_alb_99}.

The strong inhomogeneity of complex networks, implicit in their degree
distribution, changes many aspects of their behavior. In the context
of our work,  new approaches for finding the communities become
relevant. While the methods for determining  cliques, $k$-cliques and motifs
\cite{wasserman_faust_1994,degenne_forse_1994} work well if the
zero-hypothesis on the 
network structure is the Erd\H os-R\'enyi random graph, methods better
suited for complex networks were developed \cite{kul_ker_man_00,eri_sim_mas_sne_03,new_gir_02,new_gir_04,rei_bor_04,rad_cas_cec_lor_par_04,pal_der_far_vic_05,for_lat_mar_04,cap_ser_cal_col_04,gui_ama_05,for_cas_07,barber_07,lan_for_ker_08,fortunato_10,saw_amu_sal_ama_09,les_lan_mah_10,lan_kiv_sar_for_10,barber_07,lam_aus_05}. The central
quantity for majority of them is the modularity measure $Q$, which is
to be made maximal. This is achieved by various optimization
algorithms.

Here we will rely on the method of describing the
 global properties of networks using the spectral theory of graphs
\cite{chung_97}.
It deals with
eigenvalues and eigenvectors of 
various matrices representing the graph structure, which are the
adjacency matrix, Laplacian and more. 
{
It was already used  for finding clusters or communities in networks
through the properties of eigenvectors corresponding to the second largest eigenvalue
\cite{newman_06,cap_ser_cal_col_04,dan_dia_duc_are_05,don_mun_04}. In
 one step it gives the best partitioning of the network into two
 modules and repeating the algorithm recursively, the communities are found.
Our approach is  different, though. It is similar in spirit to the
analysis of covariance matrices in finance
 \cite{ple_gop_ros_am_sta_99,ple_gop_ros_ama_guh_sta_02}, where
 economic sectors are attributed to eigenvectors corresponding to the second,
the third, etc. largest eigenvalue. 
}

{
}

The first level of understanding  spectral properties of a random
matrix comprises the knowledge of the density of eigenvalues. The
second involves the localization properties of the eigenvectors. It is
the latter that is central for our approach.

Let us say first a few words on the eigenvalue density. 
Spectra of ``classical'' random graphs, like the Erd\H os-R\'enyi
model, are closely related to ``classical'' models of random matrices
\cite{mehta_91}. The typical shape of the eigenvalue density is the
Wigner semi-circle with sharp edges, with the largest eigenvalue split far off
from the bulk of all other eigenvalues. The first complication arising
in the spectrum of a random graph is the sparseness of the adjacency
matrix, which leads to the emergence of Lifschitz tails.
 This appears already in the  Erd\H os-R\'enyi
model. 
{
Despite considerable effort \cite{rod_bra_88}, the Lifschitz
tail in ER graph is still not known in all details.
}
 Asymptotic formula was
obtained by several approaches, showing that the density of states is
non-zero at arbitrarily large eigenvalues and it decays faster than
any power law \cite{rod_bra_88,sem_cug_02}.

It was soon realized that complex networks, characterized by
 power-law degree distributions, have also 
non-standard spectral properties 
\cite{far_der_bar_vic_01,goh_kah_kim_01b,dor_gol_men_02,dor_gol_men_sam_03,eri_sim_mas_sne_03,noh_rie_03,kim_hon_cho_03,far_der_jeo_etal_03,kam_chr_04,can_cap_cal_05,sim_eri_mas_sne_04,dan_dia_duc_are_05,bur_cas_05,zha_yan_wan_05,don_mun_05,sla_zha_05}.
First, there is a cusp in the middle part of the density of
eigenvalues, and second, perhaps more importantly, the tail of the
eigenvalue density seems to be described by a power law
\cite{far_der_bar_vic_01,goh_kah_kim_01b}. Numerical diagonalization
on toy models \cite{dor_gol_men_02,dor_gol_men_sam_03} as well as some
analytical estimates  confirmed  power-law
tails in the density of states. 
The replica trick \cite{rod_aus_kah_kim_05,nag_rod_08}, as well as the
cavity method \cite{dor_gol_men_sam_03,rod_cas_kuh_tak_08} 
were later adapted for scale
free networks.  It was found that
the spectrum has a power-law tail characterized by the exponent
$2\gamma-1$ related to the degree exponent $\gamma$ of the
network. Further improvements of the method were introduced recently
\cite{kuhn_08,bianconi_08}.

As we shall see, our method is similar to those used in the study of
covariance matrices of stock-market fluctuations
\cite{ga_bou_po_98,la_ci_bou_po_99,ple_gop_ros_am_sta_99,ple_gop_ros_ama_sta_00,ple_gop_ros_ama_guh_sta_02,tib_one_sar_kas_ker_06,aus_lam_07,hei_tib_sar_kas_ker_08,bou_pot_09}.
 They are
modeled as random matrices of the form $M\,M^T$, where $M$ is a random
rectangular matrix. The density of states has the Mar\v cenko-Pastur
form \cite{mar_pas_67} with sharp edges, which are smeared out into
Lifschitz tails if the matrix is sparse \cite{nag_tan_07}.

Most attention was paid to the states in the tail, i. e. located
beyond the edge of the Mar\v{c}enko-Pastur density and below the
maximum eigenvalue, which is always  
split off. These states are supposed to carry the non-trivial
information about the stock market and, indeed, the shape of the
corresponding eigenvectors was used to identify business sectors. It
was supposed that the eigenvectors were localized on  
items within specific sector
\cite{gop_ros_ple_sta_01,ple_gop_ros_ama_guh_sta_02,kim_jeo_05}.
More sophisticated 
approaches were also developed \cite{tib_one_sar_kas_ker_06}.

{
Our method owes largely to the spectral
analysis of covariance matrices. 
However, we improve these approaches by  systematic use of the
quantitative measure of localization of the eigenvectors, which is the
inverse participation ratio.  In an intuitive manner, 
similar approach was already used 
 in the analysis of gene coexpression data \cite{jal_sol_vat_li_10}.
Within this approach, we do not aim at factoring the entire network
into some number of modules, or communities, which may or may not be
overlapping, but in any case covering, as an ensemble, the whole
network. Instead, we want to find small parts of the network which
differ structurally from the rest. We may also describe our approach
as ``contrast coloring'' of the network, which makes certain
relevant parts visible against the irrelevant background. 
}

\subsection{Spectral analysis of matrices encoding the structure}

Our analysis will be devoted to bipartite graphs. There are two types
of nodes, making up sets $\mathcal{R}$ and $\mathcal{I}$. Anticipating our
application to the Amazon.com network, we think of members of
$\mathcal{R}$ as reviewers and members of $\mathcal{I}$ as items to be
reviewed. All information on the network structure is contained in the
adjacency matrix
$M$ with elements $M_{ri}\in\{0,1\}$.
 The out-degree of node $r\in\mathcal{R}$ is $k_r=\sum_i
 M_{ri}$, the in-degree of the node $i\in\mathcal{I}$ is $k_i=\sum_r
 M_{ri}$. 

In bipartite graph, the spectral properties are deduced from the
contracted matrices $B=M\cdot M^T$ and $C=M^T\cdot M$. The
interpretation of these matrices is obvious; e. g.  $B_{rs}$ 
tells us how many neighbors the nodes $r$ and $s$
 have in common.
Similar construction is used frequently in bipartite networks. As an
example, let us cite the network of tag co-occurrence in the analysis of
 collaborative tagging systems
 \cite{cat_sch_bal_ser_lor_hot_gra_stu_07} or recommendation
 algorithms investigated in \cite{zho_ren_med_zha_07}.

In order to partially separate the effects of the network structure from
the
 influence of degree distribution, we rescale the matrix elements
by the product of square 
roots of the out-degrees. This way, we get the matrix
\begin{equation}
A_{rs}=\frac{B_{rs}}{\sqrt{k_r\,k_s}}
\label{eq:ABrescaling}
\end{equation}
{with all diagonal elements equal to $1$.} We can also be more
explicit and write 
$A_{rs}=\big(\sum_i M_{ri}\,M_{si}\big)
/\sqrt{\big(\sum_i M_{ri}\big)\big(\sum_i M_{si}\big)}$. Obviously,
the matrix $A$ is symmetric.

The matrix $A$ is then diagonalized. Let us see what information
can be extracted from the eigenvalues and eigenvectors. 
First, for any square $N\times N$ matrix $D$ encoding the structure of
a graph we can interpret the traces $\frac{1}{N}\mathrm{Tr}D^k$ as
density of circles of length $k$. This number is equal to the $k$-th
moment of the density of eigenvalues of $D$. In our case, the role of
$D$ is assumed by the contracted matrix $B$ and the $k$-th moment of
$B$ expresses the density of cycles of length $2k$ on the bipartite
graph. If we use the matrix $A$ instead, the moments of the spectrum
are related to the density of weighted cycles. Each time the cycle
goes through the  vertex $r\in\mathcal{R}$, it assumes the weight
$1/k_r$. Therefore, cycles connecting vertexes with large degree are
counted with lower weight. This is just what we want here: to put
accent on  peripheral, less-connected areas of the network, rather
than on the hubs. {If we did not rescale the matrix as in
(\ref{eq:ABrescaling}), the weight of the hubs, or strongly-connected
nodes in general, would overshadow the major part of the network, where
the small communities may lie hidden.}

We expect that the spectrum has power-law tail. Indeed, it will be
confirmed in the specific example of Amazon.com, which we shall show
later. The power-law tail implies that the density of cycles beyond
certain length diverges. In terms of the limit $N\to\infty$ it means
that the number of such cycles increases faster than linearly with
$N$. The exponent of the power-law tail tells us what is the
threshold for the cycle length, beyond which the cycles are
anomalously frequent compared to the Erd\H{o}s-R\'enyi graph. 

What does all this mean for the problem of finding  small compact
communities? If we for example use the method of cliques or
$k$-cliques, we tacitly assume that the ``background'' network does
not contain many of these cliques by pure chance. But if, for example, the
tail of the spectrum of $D$ has exponent $4$, the third moment
diverges, which means that there are extremely many triangles. No
triangle, or community of size $3$,
 can therefore be considered as informationally relevant. 
That is why we consider the information on the spectrum of the network
an important auxiliary information. 

The new algorithm we propose for finding  small compact communities
relies on the properties of the eigenvectors. 
Let us denote
$e_{\lambda\,r}$ the $r$-th element of the eigenvector of the matrix $A$,
corresponding to the eigenvalue $\lambda$. To study the
localization, we need to calculate the inverse participation ratio (IPR)
defined as
\begin{equation}
q^{-1}(\lambda)=\sum_{r=1}^N (e_{\lambda\,r})^4
\end{equation}
where normalization $\sum_{r=1}^N (e_{\lambda\,r})^2=1$ is assumed.

While IPR says quantitatively to which extent an eigenvector is
localized,  this information alone is not sufficient, if we want to
draw the distinction between localized and extended states. First, it
makes no sense to ask, if a particular vector is localized, as opposed
to extended, or not. What
does make sense, however, is the question whether the states around certain
eigenvalue are localized. The way to establish that fact is by
finite-size analysis. Indeed, if $N$ is the dimension of the vector
space we work with, then 
\begin{equation}
q^{-1}(\lambda)\sim\left\{
\begin{array}{lll}
O(1),&\;N\to\infty&\text{~~localized state}\\[2mm]
O\Big(N^{-1}\Big),&\;N\to\infty&\text{~~extended state}\;.
\end{array}
\right.
\end{equation}

Second, also the shape of the density of eigenvalues changes with the
system size. When we increase $N$, the spectrum broadens. In the
textbook example of Erd\H os-R\'enyi graph, the spectrum has sharp
band edges. The edge of ER spectrum
moves as $N^{1/2}$ when $N$
grows and if  we compare the IPR at different system sizes, we must
measure the eigenvalues relative to the band edge. 
So, to compare the behavior at different sizes, we take random subset of
the network, containing $N_\mathrm{sub}$ nodes. Typically, we choose
$N_\mathrm{sub}=N/2$.
 Then, we plot the
density $D(\lambda)$ of
eigenvalues for both original network  and the density
$D_\mathrm{sub}(\lambda)$ for the random
subset. The densities are rescaled by the factor $s$, the value of
which is found empirically
 so that the data for $D(\lambda)$
and $D_\mathrm{sub}(1+(\lambda-1)s)$ overlap as much as possible. The
form of this rescaling involves the shift of the eigenvalues by $1$,
because the matrix $A$ has spectrum centered around the value
$\lambda=A_{rr}=1$. With $s$ found, we plot the IPR for the network
and the subset, with the same rescaling as used for the eigenvalues
density. The regions, where we observe that $q^{-1}(\lambda)$ remains
roughly the same for the network and its subset, are the candidate
areas where the localized states are to be found.

We continue the procedure by determining the eigenvectors with
largest $q^{-1}$ within the areas of localized states. The elements of
these vectors tell us what nodes of the network belong to the small
community. To this end, we fix a threshold $T$ and  retain only
those nodes $r\in\mathcal{R}$ for which the elements exceed the
threshold in absolute value, $|e_{\lambda\,r}|>T$. We do not propose
any exact method for fixing $T$. For the sake of consistency,
$T$ must be chosen so that the number of nodes retained is roughly
$1/q^{-1}$. In practical applications we observed the number of
retained nodes when $T$ was gradually decreased from $T=1$. At certain crossover
value of $T$ we saw that the number of nodes suddenly started increasing
substantially to much larger values than $1/q^{-1}$. So, we fixed $T$
somewhat below this crossover. We believe that this procedure could be
made automatic by a software implementation, but we did not do that.

Let us make an important remark at this point. 
Clearly, we can find some localized states also in a randomized
version of the network. These states are results of pure chance and do
not bear significant information. Therefore, we cannot exclude that
also in the true empirical network, some of the localized states
occur just accidentally and thus some of the clusters found are
spurious.
 The choice of the threshold $T$ only cannot discriminate between the
 true and the spurious clusters. However, looking at the dependence of
 IPR on eigenvalue
for the true and the randomized network
 (as will be seen later in 
Fig. \ref{fig:ec-amazon-inverse-participation-ratio-1e4-and-permuted})
we can see the regions where IPR is large and differs markedly between
true and randomized networks. The localized states found in these
regions (in
Fig. \ref{fig:ec-amazon-inverse-participation-ratio-1e4-and-permuted}
it is near the lower edge of the density of states) correspond to
clusters that are
non-random and do bear relevant information.

This way we find those vertexes $r\in\mathcal{R}$, which form the
community $\mathcal{C}_\mathcal{R}=\{r\in\mathcal{R}:|e_{\lambda\,r}|>T\}$. Next,
 we proceed by finding those $i\in\mathcal{I}$ which are
connected to them. Here we can distinguish two levels. First, a vertex
$i\in\mathcal{I}$ can be connected to at least two different vertexes from
$\mathcal{C}_\mathcal{R}$. Then, we say that it belongs to the connectors of
the community,
$\mathcal{C}_\mathcal{I}^\mathrm{con}=\{i\in\mathcal{I}:\exists
r,s\in\mathcal{C}_\mathcal{R}:r\ne s \land M_{si}=M_{ri}=1\}$. 
Further, those $i\in\mathcal{I}$
which are connected to just one vertex of $\mathcal{C}_\mathcal{R}$
form a more weakly bound part of the community, which we call cloud, 
$\mathcal{C}_\mathcal{I}^\mathrm{cloud}=\{i\in\mathcal{I}:\exists
r\in\mathcal{C}_\mathcal{R}\setminus\mathcal{C}_\mathcal{I}^\mathrm{con}:
 \land\, M_{ri}=1\}$. 
We can explicitly see the asymmetry in constructing the
community. This is due to the fact that we focused on the
diagonalization of the contraction matrix acting in the space
$\mathcal{R}$. The procedure can be, of course, performed also in the
opposite direction, diagonalizing the contraction on
$\mathcal{I}$. Both ways are equally justified on
the  formal level. The choice
should be dictated by practical reasons and by the interpretation we
want to draw from the data in any specific application.

To sum up, our procedure for finding  small communities in
bipartite networks consists in the following steps.\\\\
1. Diagonalize the matrix $A$, $A_{rs}=\big(\sum_i M_{ri}\,M_{si}\big)
/\sqrt{\big(\sum_i M_{ri}\big)\big(\sum_i M_{si}\big)}$. The output is
the density of states $D(\lambda)$ and the inverse participation
ratio $q^{-1}(\lambda)$.\\\\
2. Do the same for random subset of the network, containing half of
the nodes, find the proper rescaling factor $s$, so that rescaled
density of states for the network and the subset coincide. By
rescaling the IPR using the same factor $s$, determine the regions, in
which localized states are to be found.\\\\
3. Within the localized region, find the eigenvectors with highest
IPR. \\\\
4. For each of the eigenvectors found, determine the threshold $T$ and
establish the set $\mathcal{C}_\mathcal{R}$ 
of nodes $r$, for which $|e_{\lambda r}|>T$. This set
  is the projection of the community to the set $\mathcal{R}$. \\\\
5. Find the connector and cloud components of the community on the side of
the set $\mathcal{I}$. \\

\section{An example: Reviewing networks on Amazon.com}

\subsection{Basic structural features}

The e-commerce site Amazon.com is one of the oldest and best known on
the WWW. It has a very rich internal structure, but the user usually
sees only a small part relevant to the service requested in
a particular moment. As already announced, we shall investigate one aspect
of the Amazon.com trading, namely the network made up of connections
between
 the items to
be sold and the reviewers who have written  reports on these items. 

This network is a bipartite graph, with items
$i=1,2,\ldots,N_\mathrm{itm}$ on one side and reviewers
$r=1,2,\ldots,N_\mathrm{rev}$ on the other side. The sets of vertexes
$\mathcal{R}$ and $\mathcal{I}$ introduced in the methodical section
above, correspond to the sets of reviewers and items, respectively.

The reviews written are edges connecting these two sets.
The structure of the network can be uniquely described by the matrix
$M$, where the element 
$M_{ri}$ equals $1$ if the reviewer $r$ wrote a review on item
$i$, and $0$ otherwise.

The data were downloaded using a very simple crawler in the period from
%
%
28 July 2005 to 27 September 2005. First, a list of
total $N_\mathrm{all}=1\,714\,512$ reviewers was downloaded; at that
time the list containing all  Amazon reviewers was accessible through the
web. (It is no more so.) The list was naturally ordered by the rank Amazon assigns to each
reviewer. 
On average, reviewers with higher rank have written more reviews,
but there are exceptions. For example, at the time of data collection,
the No. 1 reviewer,
 Harriet Klausner, had written  $9581$ reviews,
while the No. 2,  Lawrance M. Bernabo, 
$10603$ reviews. This suggests that it is not only quantity but also
quality which counts when Amazon ranks their reviewers. We do not
touch here the obvious question how the most prolific reviewers do
manage reading and reviewing several books per day, throughout many
years.  As we investigate only structural features here,
these problems are left aside.

In the next step, we went through the reviewers' list, from the top rank
downwards. We looked only at about $10^5$ first reviewers and stopped
there, as we considered the sample sufficiently representative. The
remaining reviewers are only occasional writers, contributing by one
or at most a few reviews. For each reviewer we found all reviews
written by her or him  
and registered the name of the item reviewed (mostly books and CD's, of course,
but in general all kinds of goods do appear) as well as some other details
about the review. In total, we examined $99\,622$ reviewers who wrote
$2\,036\,091$  
 reviews on $645\,056$ items. 
 
\subsection{Degree distributions}

\begin{figure}[t]
\includegraphics[scale=0.95]{%
\slaninafigdir/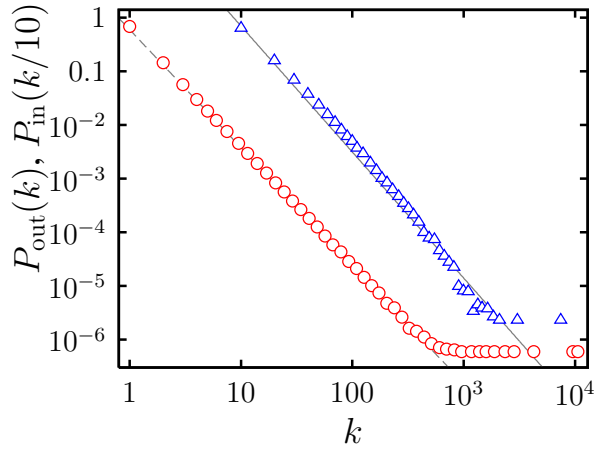}
\caption{
Degree distribution of the bipartite reviewer-product network on
Amazon.com. Circles indicate the data for out-degree (reviews per reviewer), 
triangles for
in-degree (reviews per item). 
The latter data are shifted rightwards by one decade for
better visibility. The lines are the power laws  $\propto k^{-2.2}$
(dashed line) and $\propto k^{-2.35}$
(solid line).}
\label{fig:ec-amazon-degree-distr}
\end{figure}

The simplest and most accessible local property of the network is
the degree distribution. In the list of reviewers we put down also the
reported number of reviews written by the particular person. We
neglected the possible error in this number due to various
inconsistencies. We believe that the random discrepancies between the
number of reviews reported and number of reviews which can actually be
found in the system do not influence the statistics in any
significant measure. We show the distribution as  out-degree
distribution in Fig. \ref{fig:ec-amazon-degree-distr}. We can observe 
 clear power-law dependence except for the few highest degrees. The
 exponent fitted is $\gamma_\mathrm{out}\simeq 2.2$.

Similarly we can extract the in-degree distribution from the list of
reviews. The statistics of the number of reviews per item is also
shown in Fig.\ref{fig:ec-amazon-degree-distr} and a power-law
dependence is found again. The corresponding exponent is now
$\gamma_\mathrm{in}\simeq 2.35$.  

The power distribution is by no means surprising, in view of the vast
literature on complex networks. The data provide a
clear check that Amazon.com also belongs to the class of networks with
power-law degree distribution.

\subsection{Distribution of eigenvalues}

Now we are in a position to calculate the contraction matrix $A$ acting on the 
set of reviewers, and diagonalize it. As an additional study, we
compare the results with randomized version of the reviewer-item
network. This way we discriminate
between the influence of the network structure and genuinely random
factors.

To this end, we reshuffle the edges in the reviewer-item graph, while
keeping the  degrees of all vertexes unchanged. The matrix $M$ is
replaced by $M^R$ and, correspondingly, the matrix $A$ is replaced by
$A^R$. Again, we can write $A^R_{rs}=\big(\sum_i M^R_{ri}\,M^R_{si}\big)
/\sqrt{k_r\,k_s}$. The only information on the network structure
retained here is the order sequence. As we showed in the last section,
it obeys a power law, so the features found in analyzing $A^R$ are
entirely due to power-law degree distribution, but without further
structural details.

\begin{figure}[t]
\includegraphics[scale=0.95]{%
\slaninafigdir/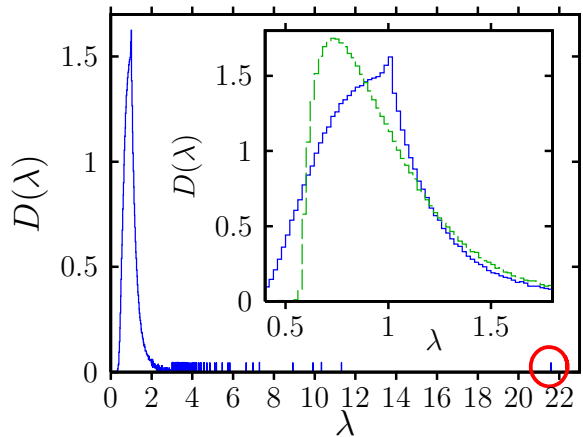}
\caption{
Distribution of eigenvalues of the reviewer-reviewer matrix. The size
of the segment is $N=10000$. For $\lambda<3$ the distribution is
plotted as a histogram, while the larger eigenvalues, $\lambda>3$ are
shown as individual vertical ticks. The largest eigenvalue is
indicated by the circle. In the inset we show the detail of the central part of
the same plot. Also in the inset, the dashed (green in color) line is
the distribution of eigenvalues of the matrix obtained by reshuffling
the reviewer-item graph. 
}
\label{fig:ec-amazon-eigenvalue-density-histogram-1e4}
\end{figure}

\begin{figure}[t]
\includegraphics[scale=0.95]{%
\slaninafigdir/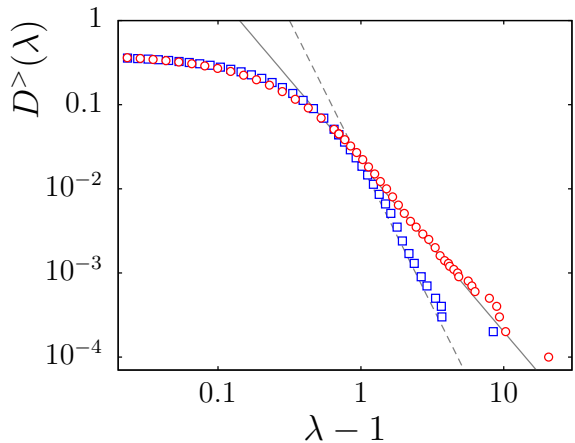}
\caption{
Detail of the upper part of the distribution of eigenvalues. The
behavior is observed using the integrated density of
eigenvalues. Circles correspond to the original reviewer-reviewer
matrix with  $N=10000$, the triangles correspond to the same matrix
subject to permutation of all its elements.
  The full line is the power $\propto
(\lambda-1)^{-2}$, the dashed line is the power $\propto
(\lambda-1)^{-3.4}$.
}
\label{fig:ec-amazon-idensity-eigenvalue-minus1}
\end{figure}

We diagonalize the matrices $A$ and $A^R$. Their eigenvalues
$\lambda_1\ge\lambda_2\ge\ldots\ge\lambda_N$
are accumulated around the value  $\lambda=1$, which corresponds to
the uniform diagonal value of both the true and the randomized matrices.  The
distributions are plotted in Figures 
\ref{fig:ec-amazon-eigenvalue-density-histogram-1e4}, 
 and
\ref{fig:ec-amazon-idensity-eigenvalue-minus1}. Let us describe now
what we can see here.

In Fig. \ref{fig:ec-amazon-eigenvalue-density-histogram-1e4} we plot
the histogram of the eigenvalues of the matrix $A$. Most
of them fall within the interval $\lambda\in[0,3]$, with sharp maximum
in the eigenvalue density at $\lambda=1$. The eigenvalues density
is much smaller
for $\lambda>3$ and we show their positions as separate
ticks. Although the notions ``bulk'' and ``tail'' are not very precise
here, we shall use them pragmatically, calling bulk the part with
$\lambda\lesssim 3$ and tail the part with $\lambda\gtrsim 3$.

In Fig.  \ref{fig:ec-amazon-eigenvalue-density-histogram-1e4} we can
also see the spectrum of the randomized matrix $A^R$.  The power-law
distribution of degrees is preserved. In the spectrum, we can observe
certain remarkable changes. 
 In the bulk of the density of states, as
shown in the inset of Fig.
\ref{fig:ec-amazon-eigenvalue-density-histogram-1e4}, the spectrum of
the reshuffled network lacks the characteristic tip at the value
$\lambda = 1$ and its shape at the lower end of the spectrum is quite
different. Most importantly, a sharp band edge develops. On the other
hand, at the upper tail of the density of states, the changes are of
minor importance.

In Fig. \ref{fig:ec-amazon-idensity-eigenvalue-minus1} we can
compare the behavior of  integrated density of states,
$D^>(\lambda)=\sum_{i,\lambda_i>\lambda} \frac{1}{N}$ in the region of
large eigenvalues.  For the
original matrix $A$ we observe a power-law decay in the tail
$D^>(\lambda)\sim(\lambda-1)^{-\tau}$ with $\tau\simeq 2$.
For the matrix $A^R$ the tail is
again quite reasonably fitted on a power law, but with larger
exponent. Let us recall that the divergence of the moments of the
eigenvalue density is related to the statistics of cycles on the
network. For the reshuffled network, 
the divergence occurs  at higher moments, therefore at 
cycle lengths longer than in the original network. This effect seems to be
a tiny one, but this is just a subtle structural difference which
goes beyond the bare degree distribution. In short, the Amazon network
has many more short loops than how many could be expected knowing only its
degree sequence. This suggests the presence of small self-reinforcing
communities. Although we do not see them yet at this stage, we can perceive
their existence through the density of states of the matrix $A$. 

Interestingly, similar conclusions about small communities were
 reached in the study of
collaborative tagging systems
\cite{cat_sch_bal_ser_lor_hot_gra_stu_07}, where two-node correlations
were calculated in order to estimate the quantity of non-randomness,
or semantic information content.

\subsection{Localization}

\begin{figure}[t]
\includegraphics[scale=0.95]{%
\slaninafigdir/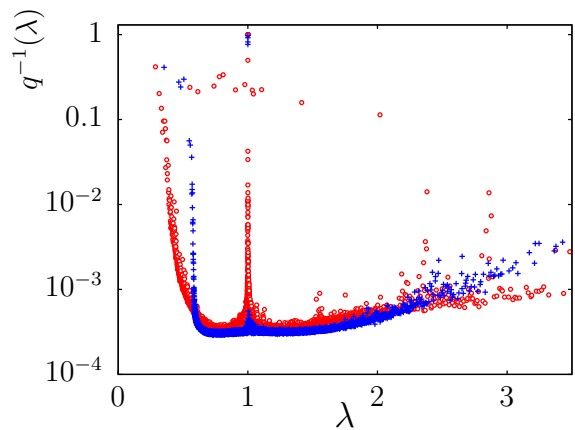}
\caption{
Inverse participation ratio for the reviewer-reviewer matrix with
$N=10000$
 ({\Large $\circ$}), and the same for
 matrix  obtained by reshuffling the reviewer-item
graph ($+$). 
Each point
denotes the IPR for the eigenvector corresponding to the indicated
eigenvalue $\lambda$.
}
\label{fig:ec-amazon-inverse-participation-ratio-1e4-and-permuted}
\end{figure}

Having investigated the eigenvalues,
let us now turn to the properties of the eigenvectors. We show in
Fig. \ref{fig:ec-amazon-inverse-participation-ratio-1e4-and-permuted} 
how the IPR depends on the eigenvalue. For the matrix $A$ we can see
larger localization around the center of the spectrum at
$\lambda=1$. Farther from the center the localization is weaker, but
it increases again at the tails, more strongly at the lower tail, while
more gradually at the upper tail. Note also some isolated highly
localized states in the bulk of the eigenvalue distribution. 

\begin{figure}[t]
\includegraphics[scale=0.95]{%
\slaninafigdir/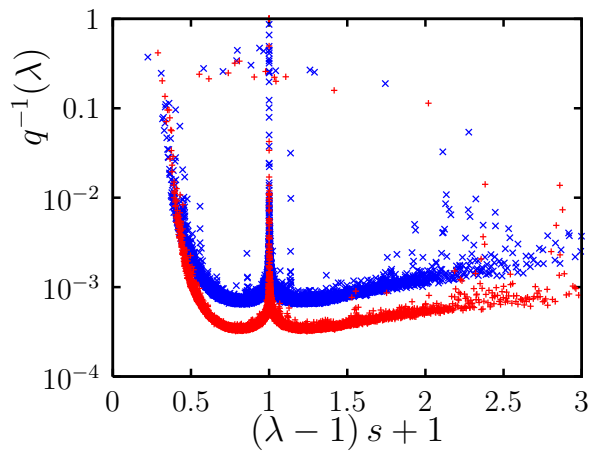}
\caption{
Inverse participation ratio for the reviewer-reviewer matrix. 
The horizontal axis is rescaled
by the  factor $s$ explained in the text. We show
the data for the matrix with 
$N=10000$
 ( $+$, $s=1$), and for the random
subset with $N_\mathrm{sub}=5000$ of the same matrix
($\times$, $s=\sqrt[3]{2}$).
Each point
denotes the IPR for the eigenvector corresponding to the indicated
eigenvalue $\lambda$.}
\label{fig:ec-amazon-ipr-scaling}
\end{figure}

Now we compare the results with the random subset of
$N_\mathrm{sub}=5000$. We found that the density of eigenvalues
coincides very well if we choose the scaling factor
$s=\sqrt[3]{2}=\sqrt[3]{N/N_\mathrm{sub}}$. With the same scaling we
plot the IPR in 
Fig. \ref{fig:ec-amazon-ipr-scaling}. We can see that the absence of a
clearcut band edge is complemented by the absence of any region of
localized states at the upper end of the spectrum. The lower end does
show localized states, though. Therefore, the candidates for 
compact communities are to be found close to the lower end of the
spectrum. In the next section we describe what we have found there.

\begin{figure}[t]
\begin{center}
\includegraphics[scale=0.5]{%
\slaninafigdir/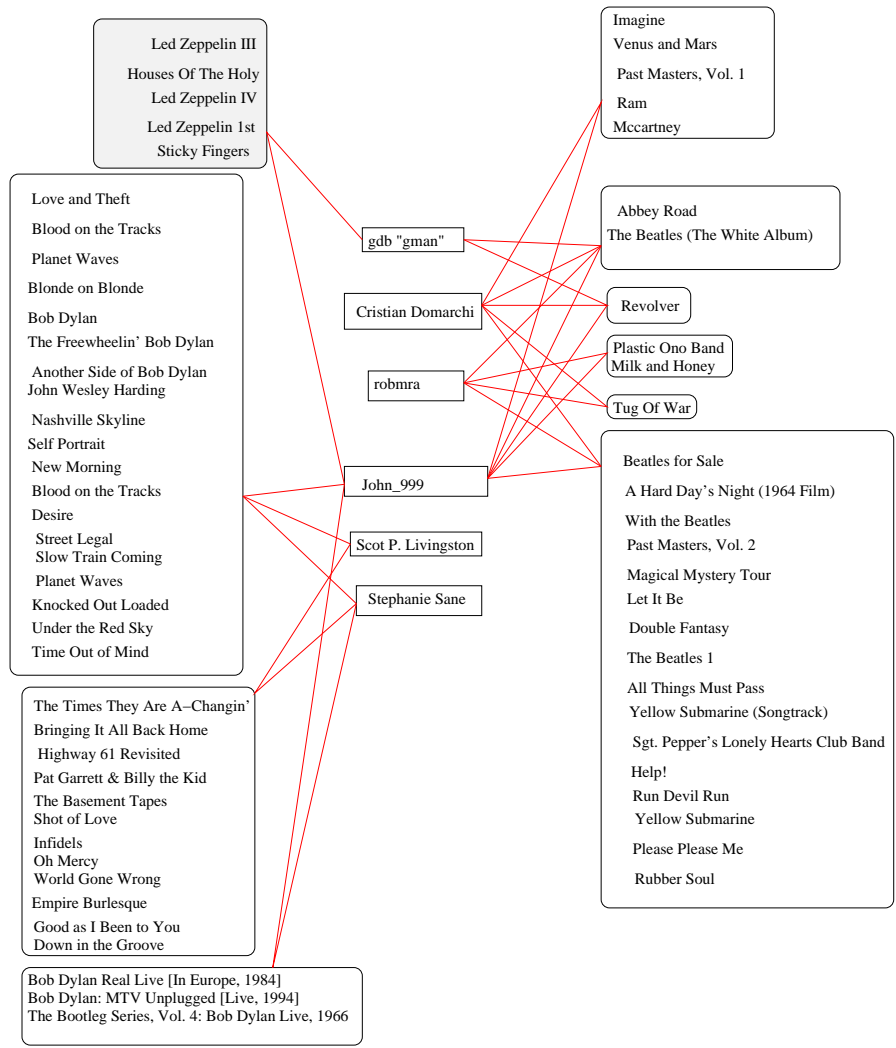}
\end{center}
\caption{
The ``pop-music'' 
community in the network producing a very localized eigenvector of
the matrix $A$. In the middle, code-names of the reviewers, on the
right, recordings by The Beatles (mostly as a band, some other by
individual members), on the left, recordings by Bob
Dylan, with exception of the shaded box which contains four times
music by Led Zeppelin and once Rolling Stones.} 
\label{fig:very-localized-cluster-5-2}
\end{figure}

\begin{figure}[t]
\begin{center}
\includegraphics[scale=0.5]{%
\slaninafigdir/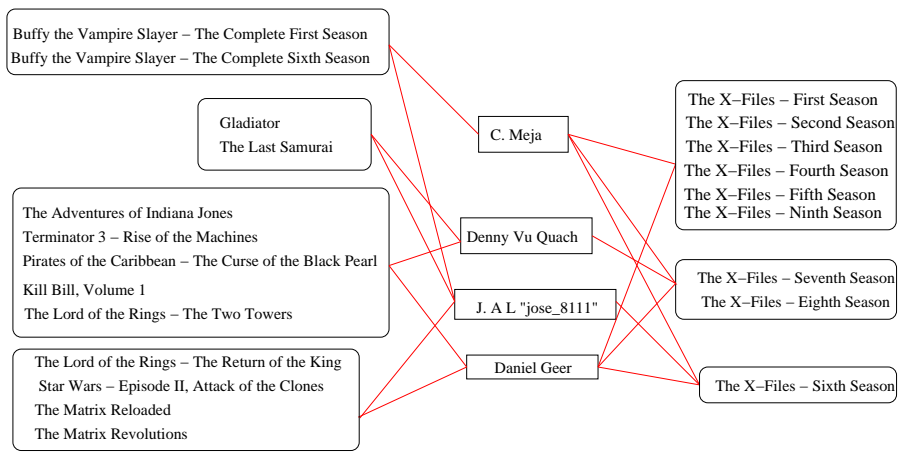}
\end{center}
\caption{
The ``pop-movie'' 
community in the network producing a very localized eigenvector of
the matrix $A$. In the middle, code-names of the reviewers, on the
right, the X-Files series, on the left, other popular movies.}
\label{fig:very-localized-cluster-6-12}
\end{figure}

\begin{figure}[t]
\begin{center}
\includegraphics[scale=0.5]{%
\slaninafigdir/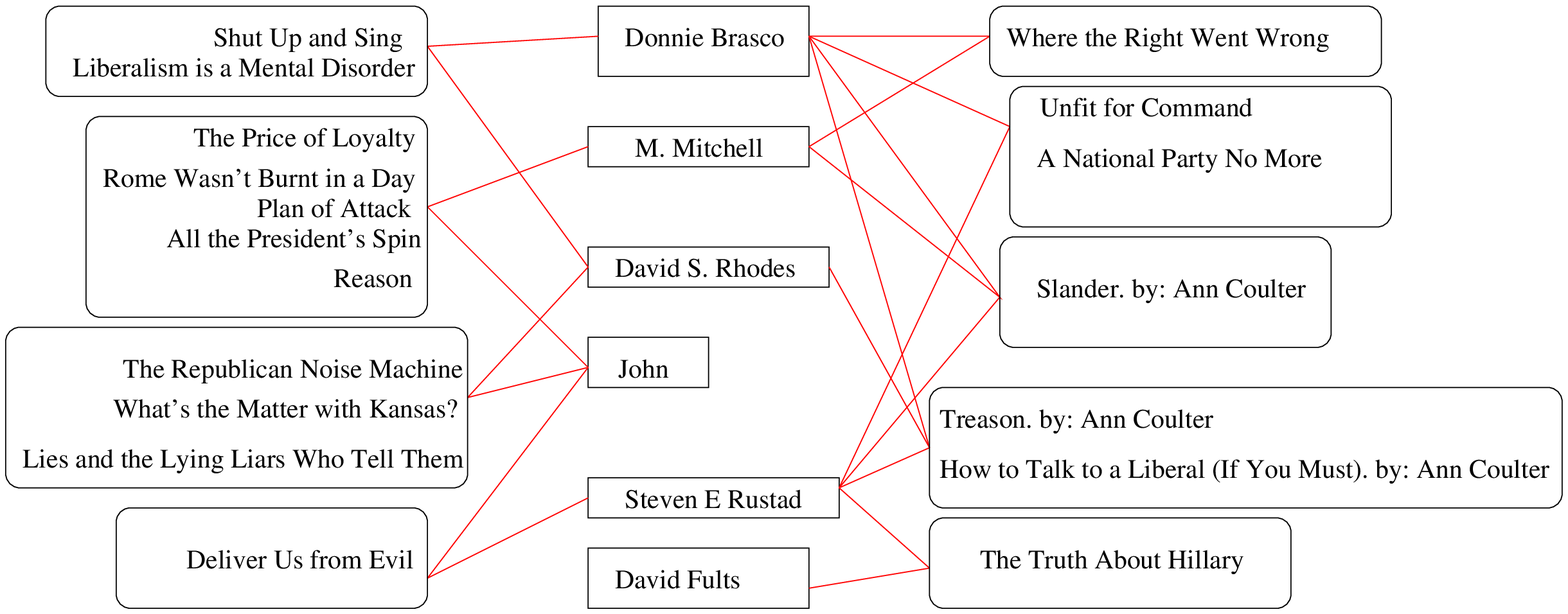}
\end{center}
\caption{
The ``pop-politics'' 
community in the network producing a very localized eigenvector of
the matrix $A$. In the middle, code-names of the reviewers, on the left and
right, books treating mainly the clash of {(neo-) conservatives} versus
liberals in the USA. Note that Ann Coulter is the most prominent
book author in this community. 
}
\label{fig:very-localized-cluster-7-14}
\end{figure}

\section{Finding and interpreting the {communities}}
As we have said, the most localized states are the candidates for small
{and densely interlinked}
 communities of reviewers. We counted as members of
 the community only those reviewers, whose element in the
 eigenvector was larger than a threshold, $|e_{\lambda\,r}|>T$. The
 value of the threshold $T$  was found by trial-and-error, so that  
all relevant
nodes, on which localization appears, were kept, while the remaining
ones, interpreted as a
noisy neighborhood, were left out. This adjustment of  thresholds
also indicates that 
the borders of the communities found in this way are not sharp.
{In our set of $10^4$
 reviewers, the number of communities which can be
considered as well-localized is about $\simeq 10$. We were able to explicitly
draw and interpret 
$7$ communities. With average size of the communities around 6 people, the
fraction of reviewers in small compact communities can be estimated to
about $0.5$ per cent. In other words, we have been able to find
relatively rare cases when fractional segments of the network display
anomalously high density of mutual links. {However, we}
 expect that this fraction would rapidly grow if more
reviewers are included from the top of the Amazon list downwards. From
this point of view  the small percentage of the reviewers in  small
communities is partly an artifact due to the choice of the reviewers
starting from the top of the list of the most productive Amazon.com reviewers.}

 {Now, let us look at several specific examples of the communities found.}
The first example of such a small grouping is shown in
Fig. \ref{fig:very-localized-cluster-5-2}. 
 (In this case we took the $5$th most localized vector,
$q^{-1}=0.095675$,
corresponding eigenvalue $\lambda=0.359$,
and the threshold was  taken as $T=0.2$.)
The items reviewed by the reviewers within the
  community found in this way are of two types. First, there are
  those reviewed by at least two reviewers from the community. These
  items keep the community together and we call them
  ``connectors''. We show them in
  Fig. \ref{fig:very-localized-cluster-5-2} linked to their
  corresponding reviewers. However, one should note that 
the reviewers themselves play the role of ``connectors'' for the
  items, to the same extent as the items are ``connectors'' for the
  reviewers. 
Second, there are items reviewed by only
  one reviewer of the community. These items form a kind of 
  ``cloud'' around the core of the network segments. We do not show the
  ``cloud'' in our figures, but we shall discuss its meaning later.

{What are the product-connectors in the given community 
(Fig. \ref{fig:very-localized-cluster-5-2})? 
We can see that the
  maximum number  
of reviewers for
one item is $4$  and it holds for two audio
recordings: ``The Beatles (The White Album)'' and ``Abbey Road''
also by Beatles.
Thus, the core of the community is kept together by one of the most
popular music bands ever. 
{The remaining items are thematically close.
They refer to other records by Beatles and also by Beatles ex-members,
or to the music of Bob Dylan. (Ex-)Beatles and Dylan 
 cover about a
half of the items each. The only exception is a small set of five
recordings of other pop-classics, namely four of  Led Zeppelin and one of Rolling
Stones. In short,  all items fall into the range of  notoriously
known pop-music
stars. It is interesting that this characteristic does not concern
the connecting items only, but majority of all other reviews by the
members of the community (not included in the graph). Thus, not only the
connectors, but also the ``cloud'' bears the same characteristics.}

{Therefore, the interest of
these reviewers lies, in general, 
 within a rather narrow scope determined by the pair
Dylan-Beatles, with some small excursions farther into mainstream 
pop-music, similar to
the small ``Led Zeppelin'' set in Fig.
\ref{fig:very-localized-cluster-5-2}. For example, the reviewer
gdb has also written on CD's by  U2 and David Bovie, while the
``cloud'' reviews by Cristian Domarchi (not listed in
Fig. \ref{fig:very-localized-cluster-5-2}) 
pertain only to other recordings by ex-Beatles plus one book;
among all the $6$ reviewers, only Stephanie Sane shows interests which go
clearly beyond the Dylan-Beatles repertoire, 
reviewing a good deal of books, mostly
mystery and detective fiction.}

A similar picture is provided by the analysis of
 other communities. Let us very briefly describe two
more of them.

The first one (Fig.
 \ref{fig:very-localized-cluster-6-12}) belongs 
  to another pop-cultural domain, this time
concentrating on 
DVD movies with a sci-fi and fantasy flavor. Again, we found that the
reviewers are active within rather narrow bounds. They focus on widely
popular titles, overlapping very little  with any  other possible themes
or genres. 
 Only a small part of the
reviews by the members of the community 
are related to something else, e.g., to
 books by M. Proust and T. Mann.

The third and last example we want to mention is shown in
Fig.  \ref{fig:very-localized-cluster-7-14}. In analogy with the
former examples, the ``pop-music'' and ``pop-movie'' communities, we may 
call
this one a ``pop-politics'' community. The reviewers here concentrate on
books discussing the presidency of G. Bush, the evils of liberal
ideology, as compared with  neo-conservatism, and so on. The
core of the community is kept together by the books of Ann Coulter, who
is known as a militant anti-liberal writer. Majority 
 of the books in this group is targeted at the widest
public, as is the music by The Beatles and movies of the ``X-Files'' type. 
Their themes are not esoteric, these products are not aiming at
specialized audiences; yet, the zeal of the
reviewers makes a ``cult'' of them. Again, this
 community is narrowly defined
by the interest in these popular issues and not much else. In the
``cloud''
 of
other items reviewed by the members of this community
 we find some other books by Ann
Coulter, accompanied by 
books such as (the titles are self-explaining, we believe) 
{\it Worse Than Watergate: The Secret Presidency of
  George W. Bush}; 
{\it Blinded by the Right:
The Conscience of an Ex-Conservative};
{\it A Matter of Character :
Inside the White House of George W. Bush}; 
{\it The Family : The Real
Story of the Bush Dynasty}; {\it Chain of Command : The
Road from 9/11 to Abu Ghraib}, and similar. Out of the six reviewers,
 only  Donnie
Brasco shows some additional field of interest, having written about
various pop-music CD's as well. 

Let us  sum up these observations (supported also by
 analyses of other small
communities we were able to find in the sample). 
Our expectations  that strongly localized eigenvectors would reveal some
specific small communities was fulfilled in the sense that we have
indeed found groups of zealots, concentrated  on a
relatively narrow segment of commodities sold on
 Amazon.com. Individual interests of these reviewers
only scarcely reach beyond the theme common to the community.

On the other hand, however, it would be misleading to imagine these people
as eccentrics focused on highly specialized, marginal or even
extreme cultural artifacts. The subjects of their reviews are quite
ordinary, clearly part of the cultural mainstream. And, by their
tastes, the reviewers themselves seem belonging to wide audiences,
often focused on classics or well-established pop-cultural
products. In other words, anomalous tiny fragments of this huge
network, characteristic by various authors repeatedly writing reviews
on the same items, refer typically not to some marginal cultural forms
with specialized contents, but rather to widely shared cultural tastes
and mainstream enthusiasts. 

A more detailed analysis of these findings is beyond the scope of this
methodological paper and its analytical illustration. Very probably,
several possible explanations could turn valid in parallel, including
the nature of the Amazon.com portal (primarily designed for general
audiences and as wide consumer population as possible), possibly
higher probability that reviews on widely favored artifacts get
``localized'' etc.
What should perhaps be stressed here, however, is the peculiar
character of the communities or network segments under
discussion. It is clear that the tiny network fragments counting 5 or
10 reviewers and dozens of reviews cannot represent ``big''
consumer populations and ``widespread'' artifacts in some
straightforward way. Rather, they may provide a rather specific
(``small-scale'') way of looking at a mass-scale phenomenon. Let
us tell something more about this specificity. 

We have already noted that the network and its segments we are
studying is not a ``social network'' as traditionally
envisaged. The interaction constituting the network is so massively
mediated and by-produced (while remaining observable, ``real''
enough and grounded in intentional social action) that we leave the
territory of what is usually counted by social scientists as a
``group'' or ``community''. But even more is at stake in this
direction. A closer view of our findings reveals that one cannot
unambiguously say whether the ``connecting'' reviewed products
provide interpretive framework for statements about the reviewers, or
whether -- on the contrary -- it is reviewers and their actions
that provide clues for interpreting communities of products. In other
words, we are unable to determine whether we study groupings of people
(connected by products) or of commodities (connected by people). In
fact, we should better try to understand both within a single hybrid network,
meaningfully connected. While studying phenomena of product reviewing,
products and reviewers cannot be separated. The sets of products
represented in our figures 
(Fig. \ref{fig:very-localized-cluster-5-2}
 \ref{fig:very-localized-cluster-6-12}
 \ref{fig:very-localized-cluster-7-14}) 
do not simply make sense
(and do not hold together) without the reviews written about them by the
represented reviewers. Indeed, the products grouped by, e.g.,
purchases carried out by Amazon.com users would look differently. On
the other hand, the groups of reviewers would not make sense without
the particular reviewed products (their amount and nature). Thus, we
believe the segments identified in our example can directly represent
neither populations of consumers nor entire sections in the Amazon.com
commodities catalog. Rather, they represent, in a complex way and as
if under a specific lens, a phenomenon of on-line user reviewing,
better understanding of which may contribute to our knowledge of
contemporary popular culture and technologically mediated economic
processes.

\section{Conclusions}

Thanks to numerous sociological efforts in the field of social network analysis as
well as the work on networks done in other scientific disciplines such
as theoretical physics, various mathematical tools have been
developed. They aim either at determining large-scale structures in
complex networks or at identification of smaller network segments such
as cliques or acquaintances. In this work, we introduced a new
mathematical procedure relatively close to the latter type of task. We
believe the method is well suitable for finding the most relevant
small segments of complex networks, when ``relevance'' cannot or need
not be equaled to some absolute level of mutual connectivity between
the nodes.  We  argue that this is often the case,
because important social forces or processes are often related to
highly mediated and heterogeneous groupings, typically constituted as
by-products of various, differently oriented actions, and where people
characteristically and usually do not intentionally address each other
and even do not know each other (here, we could speak of ``ultra-weak'' ties).
The proposed method based on well-localized eigenvectors is well
capable to find these small communities with anomalously high density of
mutual links and therefore reveal a kind of semantic information
hidden in the network, otherwise often neglected. As such, our method
may be a good starting point for more fine-grained further analysis of
given phenomena.

As an empirical example, we have chosen the data available from the
Amazon.com on-line shopping portal. We studied the network constituted
by users writing reviews of the same products offered for purchase on
the website during the summer 2005, when the data were gathered. 
Reviewers become connected if they
have written a review on an identical item. When such connections
locally proliferate we get a grouping of relevance.

These groupings are not directly related to the
top-lists of popularity, but reveal the most focused points in the
network. They are constituted by socially rather distant ties, i.e.,
by a kind of ultra-weak ties, namely highly mediated links by-produced
during processes primarily aimed at something else than addressing
each other to establish acquaintance or become closer.

The first important result of our analysis 
 is the power-law tail in the density of
eigenvalues. This feature is partially, but not entirely, due to the
power-law degree distribution. Comparing the spectrum arising
from the network with the spectrum of a random network with the same
degree sequence, we find a power-law tail in both cases, but the
exponent is significantly smaller in the original network. 
Generally, such a tail implies that the density of cycles beyond
certain length diverges when the size of the network tends to
infinity. The difference in the exponent means that some shorter
cycles keep finite density in the randomized network, while in the
original one they are much more abundant. This means that the Amazon network
contains much more compact groupings 
than what would be expected
knowing only its degree sequence.

To see at least some of these small groupings, we looked at
 well-localized eigenvectors. These
localized states represent small communities {or network
 segments}
 and bear  semantic
information hidden in the network. {We call them ``hot
 spots'', as they represent local structures which differ from the
 surrounding background.}
We were able to explicitly find some of these communities and
 {attribute   meaning to them}. The three of them 
{briefly discussed} in this paper can
 be labeled as pop-music, pop-movie, and pop-politics communities. The
 reviewers of these communities are very strongly focused on one
 narrow segment. This segment itself belongs usually
 {to mass or popular} 
 culture, so {it cannot be considered as marginal or
 esoteric}. It is the enthusiasm of 
 the reviewers which singles the segment out of the sea of millions
 products traded on Amazon.com.

 Our analysis shows that {only} about half
per cent of the reviewers belong to these network segments in the small
sample of $10^4$ reviewers. {However, we}
 expect that this fraction would rapidly grow if more
reviewers are included from the top of the Amazon list downwards. 
{If carefully treated and interpreted the identified
  network segments may be useful for enhancing our knowledge of mass
  or popular culture and complex economic processes related to
  E-consumerism.} Especially, it would be interesting to make
systematic classification of the small communities.

Besides these specific findings we would like to highlight another, more
general feature. When analyzing the chosen example, it turned out that
conventional talking about ``networks of reviewers'' might be
sociologically misleading. Our groupings, in fact, were constituted
not only by people writing reviews on the same products, but also
(simultaneously) by products reviewed by the same reviewers. That is
why we decided to switch to a more appropriate
 term ``networks of reviewing''. This
term indicates the hybrid nature of networks we have been dealing with
and it allows better talking about processes of online economy rather than on
bare structures composed of its human agents. In this respect, our
approach is well compatible with the currently increasing emphasis
on heterogeneity as an essential quality of collectivities studied by
social scientists \cite{latour_05}.

The method can be applied in a straightforward way to any kind of
network, whereever the data can be collected easily. However, technical
limitations of the method may arise in networks larger that several
tens of thousands of vertices, {due to computer memory
  limitations}. As shown also by the example of Amazon.com,
 on-line networks are often larger than that. 
Then, we must decide which subset
of the whole network can be considered representative. In our case we
chose
 the subset of the most productive reviewers, but other networks might
require other criteria.

\section*{Acknowledgments}
FS wishes to thank Y.-C. Zhang for encouragement and inspiring comments
and to Fribourg University for support and hospitality during the stay
when this study was started.  
This work was carried our within the project AV0Z10100520 of the Academy 
of Sciences of the Czech republic and was  
supported by the M\v{S}MT of the Czech Republic, grant no. 
OC09078 and by the Research Program CTS MSM 0021620845.


\begin{thebibliography}{99}
\bibitem{alb_bar_01}
 Albert, R. and  Barab\'asi, A.-L., 
Statistical mechanics of complex networks, 
{\it Rev. Mod. Phys.},
 {\bf 74} 
 (2002) 
 47-97.

%
%
\bibitem{allen_2002}
Allen, J.,  Symbolic economies: The "culturalization" of economic knowledge, in: {\it Cultural economy: Cultural analysis and commercial life},  eds. P. d. Gay and M. Pryke (Sage, London, Thousand Oaks and New Delhi, 2002) pp. 39-58. 


\bibitem{aus_lam_07}
 Ausloos, M. and  Lambiotte, R., 
Clusters or networks of economies? A macroeconomy study through Gross Domestic Product, 
{\it Physica A},
 {\bf 382} 
 (2007) 
 16-21.

\bibitem{bar_alb_99}
 Barab\'asi, A.-L. and  Albert, R., 
Emergence of Scaling in Random Networks, 
{\it Science},
 {\bf 286} 
 (1999) 
 509-512.

\bibitem{barber_07}
 Barber, M. J., 
Modularity and community detection in bipartite networks, 
{\it Phys. Rev. E},
 {\bf 76} 
 (2007) 
 066102.

%
%
\bibitem{barry_slater_2005}
Barry, A. and Slater, D., eds.,  {\it The technological economy} (Routledge, London and New York 2005).


%
%
\bibitem{beer_2008}
Beer, D.,  Making friends with Jarvis Cocker: Music culture in the context of Web 2.0, {\it Cultural Sociology}, {\bf 2} (2008)  222-241.


%
%
\bibitem{benkler_2006}
Benkler, Y.,  {\it The wealth of networks: How social production transforms markets and freedom} (Yale University Press, New Haven and London, 2006).


%
%
\bibitem{berger_luckmann_1967}
Berger, P. L. and Luckmann, T.,  {\it The social construction of reality: A treatise in the sociology of knowledge} (Penguine Books, Harmondsworth, 1967).


\bibitem{bianconi_08}
 Bianconi, G., 
Spectral properties of complex networks, 
{\it arXiv:0804.1744}
 (2008).

\bibitem{bollobas_85}
 Bollob\'as, B., 
{\it Random Graphs}
 (Academic Press, London, 1985). 

\bibitem{bor_sch_03}
 Bornholdt, S. and  Schuster , H. G., 
{\it Handbook of Graphs and Networks}
 (Wiley-VCH, Weinheim, 2003). 

\bibitem{bou_pot_09}
 Bouchaud, J.-P. and  Potters, M., 
Financial Applications of Random Matrix Theory: a short review, 
{\it arXiv:0910.1205}, to appear in: {\it Handbook of Random Matrix Theory}.


\bibitem{bur_cas_05}
 Burioni, R. and  Cassi, D., 
Random walks on graphs: Ideas, techniques and results, 
{\it J. Phys. A.},
 {\bf 38} 
 (2005) 
 R45.

%
%
\bibitem{callon_etal_2005}
Callon, M., M\'eadel, C., and Rabeharisoa, V.,  The economy of qualities, in: {\it The technological economy}, eds. A. Barry and D. Slater  (Routledge, London and New York, 2005) pp. 28-50.


\bibitem{cap_ser_cal_col_04}
 Capocci, A.,  Servedio,  V. D. P.,  Caldarelli,  G. and  Colaiori, F., 
Detecting communities in large networks, 
{\it Physica A},
 {\bf 352} 
 (2005) 
 669-676.

%
%
\bibitem{castells_2000}
Castells, M.,  Toward a sociology of the network society, {\it Contemporary Sociology} {\bf 29 }  (2000) 693-699.


\bibitem{cat_sch_bal_ser_lor_hot_gra_stu_07}
 Cattuto, C.,  Schmitz,  C.,  Baldassarri,  A.,  Servedio,  V. D. P.,  Loreto,  V.,  Hotho,  A.,  Grahl,  M. and  Stumme,  G., 
Network Properties of Folksonomies, 
{\it AI Communications},
 {\bf 20} 
 (2007) 
 245-262.

%
%
\bibitem{christakis_fowler_2009}
Christakis, N. A. and Fowler, J. H.,  {\it Connected: The surprising power of our social networks and how they shape our lives} (Hachette Book Group, New York, 2009).


\bibitem{chung_97}
 Chung, F. R. K., 
{\it Spectral Graph Theory}
 (American Mathematical Society, 1997). 

\bibitem{dan_dia_duc_are_05}
 Danon, L.,  Duch,  J.,  Arenas,  A. and  Diaz-Guilera, A., 
Comparing community structure identification, 
{\it J. Stat. Mech},
 (2005) 
 P09008.

%
%
\bibitem{david_pinch_2008}
David, S. and Pinch, T.,  Six degrees of reputation: The use and abuse of online review and recommendation systems, in:  {\it Living in a material world: Economic sociology meets science and technology studies} eds. T. Pinch and R. Swedberg (The MIT Press, Cambridge, MA and London,  2008) pp. 341-374.


%
%
\bibitem{degenne_forse_1994}
Degenne, A. and Forse, M.,  {\it Introducing social networks}  (Sage, London, Thousand Oaks and New Delhi, 1994).


\bibitem{diestel_00}
 Diestel, R., 
{\it Graph Theory}.
 (Springer, New York, 2000). 

\bibitem{don_mun_04}
 Donetti, L. and  Mu\~{n}oz, M. A., 
Detecting network communities: a new systematic and efficient algorithm, 
{\it J. Stat. Mech.},
 (2004) 
 P10012.

\bibitem{don_mun_05}
 Donetti, L. and  Mu\~{n}oz, M. A., 
Improved spectral algorithm for the detection of network communities, 
{\it physics/0504059},
 (2005). 

\bibitem{dor_men_01}
 Dorogovtsev, S. N. and  Mendes, J. F. F., 
Evolution of random networks, 
{\it Adv. Phys.},
 {\bf 51} 
 (2002) 
 1079-1187.

\bibitem{dor_men_03}
 Dorogovtsev, S. N. and  Mendes, J. F. F., 
{\it Evolution of Networks}.
 (Oxford University Press, Oxford, 2003). 

\bibitem{dor_gol_men_02}
 Dorogovtsev, S. N.,  Goltsev,  A. V. and  Mendes, J. F. F., 
Pseudofractal Scale-free Web, 
{\it Phys. Rev. E},
 {\bf 65} 
 (2002) 
 066122.

\bibitem{dor_gol_men_sam_03}
 Dorogovtsev, S. N.,  Goltsev,  A. V.,  Mendes,  J. F. F. and  Samukhin, A. N., 
Spectra of complex networks, 
{\it Phys. Rev. E},
 {\bf 68} 
 (2003) 
 046109.

%
%
\bibitem{emirbayer_goodwin_1994}
Emirbayer, M. and Goodwin, J.,  Network analysis, culture, and the problem of agency, {\it American Journal of Sociology} {\bf 99 }  (1994) 1411-1453.


\bibitem{erd_ren_59}
 Erd\H os, P. and  R\'enyi, A., 
On random graphs I, 
{\it Pub. Math. Debrecen},
 {\bf 5} 
 (1959) 
 290-297.

\bibitem{eri_sim_mas_sne_03}
 Eriksen, K. A.,  Simonsen,  I.,  Maslov,  S. and  Sneppen, K., 
Modularity and Extreme Edges of the Internet, 
{\it Phys. Rev. Lett.},
 {\bf 90} 
 (2003) 
 148701.

\bibitem{far_der_bar_vic_01}
 Farkas, I. J.,  Der\'enyi,  I.,  Barab\'asi,  A.-L.,  Vicsek,  T., 
Spectra of ``Real-World'' graphs: Beyond the semi-circle law, 
{\it Phys. Rev. E},
 {\bf 64} 
 (2001) 
 026704.

\bibitem{far_der_jeo_etal_03}
 Farkas, I.,  Der\'enyi,  I.,  Jeong,  H.,  N\'eda,  Z.,  Oltvai,  Z. N.,  Ravasz,  E.,  Schubert,  A.,  Barab\'asi,  A.-L. and  Vicsek, T., 
Networks in life: Scaling properties and eigenvalue spectra, 
{\it Physica A},
 {\bf 314} 
 (2002) 
 25-34.


\bibitem{can_cap_cal_05}
 Ferrer i Cancho, R.,  Capocci,  A. and  Caldarelli, G., 
Spectral methods cluster words of the same class in a syntactic dependency network, 
{\it cond-mat/0504165},
 (2005). 

%
%
\bibitem{foot_2009}
Foot, K.,  Internet use and meaning, materiality, and sociality through activity theory. Presentation at the 4S Annual Conference, Washington D.C., Oct. 31, 2009 (2009).


\bibitem{fortunato_10}
 Fortunato, S., 
Community detection in graphs, 
{\it Phys. Rep.},
 {\bf 486} 
 (2010) 
 75-174.

\bibitem{for_cas_07}
 Fortunato, S. and  Castellano, C., 
Community Structure in Graphs, 
 in: {\it Encyclopedia of Complexity and System Science},
 (Springer, New York, 2009).


\bibitem{for_lat_mar_04}
 Fortunato, S.,  Latora,  V. and  Marchiori, M., 
Method to find community structures based on information centrality, 
{\it Phys. Rev. E},
 {\bf 70} 
 (2004) 
 056104.

\bibitem{ga_bou_po_98}
 Galluccio, S.,  Bouchaud,  J.-P. and  Potters, M., 
Rational decisions, random matrices and spin glasses, 
{\it Physica A},
 {\bf 259} 
 (1998) 
 449-456.

\bibitem{goh_kah_kim_01b}
 Goh, K.-I.,  Kahng,  B. and  Kim, D., 
Spectra and eigenvectors of scale-free networks, 
{\it Phys. Rev. E},
 {\bf 64} 
 (2001) 
 051903.

\bibitem{gop_ros_ple_sta_01}
 Gopikrishnan, P.,  Rosenow,  B.,  Plerou,  V. and  Stanley, H. E., 
Quantifying and interpreting collective behavior in financial markets, 
{\it Phys. Rev. E},
 {\bf 64} 
 (2001) 
 035106.

%
%
\bibitem{granovetter_1983}
Granovetter, M.,  The strength of weak ties: A network theory revisited {\it Sociological Theory} {\bf 1 }  (1983) 201-233.


\bibitem{granovetter_73}
 Granovetter, M., 
The Strength of Weak Ties, 
{\it American Journal of Sociology},
 {\bf 78} 
 (1973) 
 1360-1380.

\bibitem{gui_ama_05}
 Guimer\`a R. and  Amaral, L. A. N., 
Functional cartography of complex metabolic networks, 
{\it Nature},
 {\bf 433} 
 (2005) 
 895-900.

%
%
\bibitem{hacking_1999}
Hacking, I.,  {\it The social construction of what? }  (Harvard University Press, Cambridge, 1999).


\bibitem{hei_tib_sar_kas_ker_08}
 Heimo, T.,  Tib\'ely,  G.,  Saram\"aki,  J.,  Kaski,  K. and  Kert\'esz, J., 
Spectral methods and cluster structure in correlation-based networks, 
{\it Physica A},
 {\bf 387} 
 (2008) 
 5930-5945.

\bibitem{jal_sol_vat_li_10}
 Jalan, S.,  Solymosi,  N.,  Vattay,  G. and  Li, B., 
Random matrix analysis of localization properties of gene coexpression network, 
{\it Phys. Rev. E},
 {\bf 81} 
 (2010) 
 046118.

\bibitem{kuhn_08}
 K\"uhn, R., 
Spectra of Sparse Random Matrices, 
{\it J. Phys. A: Math. Theor.},
 {\bf 41} 
 (2008) 
 295002. 

\bibitem{kam_chr_04}
 Kamp, C. and  Christensen, K., 
Spectral analysis of protein-protein interactions in Drosophila melanogaster, 
{\it Phys. Rev. E},
 {\bf 71} 
 (2005) 
 041911.

\bibitem{kim_jeo_05}
 Kim, D.-H. and  Jeong, H., 
Systematic analysis of group identification in stock markets, 
{\it Phys. Rev E},
 {\bf 72} 
 (2005) 
 046133.

\bibitem{kim_hon_cho_03}
 Kim, B. J.,  Hong,  H. and  Choi, M. Y., 
Netons: Vibrations of complex networks, 
{\it J. Phys. A: Math. Gen.},
 {\bf 36} 
 (2003) 
 6329. 

\bibitem{kul_ker_man_00}
 Kullmann, L.,  Kert\'esz,  J. and  Mantegna, R. N., 
Identification of clusters of companies in stock indices via Potts super-paramagnetic transitions, 
{\it Physica A},
 {\bf 287} 
 (2000) 
 412-419.

\bibitem{la_ci_bou_po_99}
 Laloux, L.,  Cizeau,  P.,  Bouchaud,  J.-P. and  Potters, M., 
Noise Dressing of Financial Correlation Matrices, 
{\it Phys. Rev. Lett.},
 {\bf 83} 
 (1999) 
 1467-1470.

\bibitem{lam_aus_05}
 Lambiotte, R. and  Ausloos, M., 
Uncovering collective listening habits and music genres in bipartite networks, 
{\it Phys. Rev. E},
 {\bf 72} 
 (2005) 
 066107.

\bibitem{lan_for_ker_08}
 Lancichinetti, A.,  Fortunato,  S. and  Kert\'esz, J., 
Detecting the overlapping and hierarchical community structure of complex networks, 
{\it New J. Phys.},
 {\bf 11} 
 (2009) 
 033015. 

\bibitem{lan_kiv_sar_for_10}
 Lancichinetti, A.,  Kivel\"a,  M.,  Saram\"aki,  J. and  Fortunato, S., 
Characterizing the community structure of complex networks, 
{\it PLoS ONE}, 
{\bf 5 }
 (2010) e11976. 

%
%
\bibitem{lash_urry_1994}
Lash, S. and Urry, J.,  {\it Economies of signs and space} (Sage, London, 1994).


\bibitem{latour_05}
 Latour, B., 
{\it Reassembling the social: An introduction to actor--network theory}
 (Oxford University Press, Oxford, 2005).

\bibitem{les_lan_mah_10}
 Leskovec, J.,  Lang,  K. J. and  Mahoney,  M. W., 
Empirical Comparison of Algorithms for Network Community Detection, 
{\it in: WWW '10: Proceedings of the 19th international conference on World wide web},
 (2010) pp. 631-640. 

%
%
\bibitem{licoppe_2008}
Licoppe, C.,  Understanding and reframing the electronic consumption experience: The interactional ambiguities of mediated coordination, in:  {\it Living in a material world: Economic sociology meets science and technology studies} eds. T. Pinch and R. Swedberg (The MIT Press Cambridge, MA and London,    2008) pp. 317-340.


\bibitem{mar_pas_67}
 Mar\v{c}enko, V. A. and  Pastur, L. A., 
Distribution of eigenvalues for some sets of random matrices, 
{\it Math. USSR--Sbornik},
 {\bf 1} 
 (1967) 
 457-483.

\bibitem{mehta_91}
 Mehta, M. L., 
{\it Random matrices}.
 (Academic Press, San Diego, 1991). 

\bibitem{nag_rod_08}
 Nagao, T. and  Rodgers, G. J., 
Spectral density of complex networks with a finite mean degree, 
{\it J. Phys. A: Math. Theor.},
 {\bf 41} 
 (2008) 
 265002.

\bibitem{nag_tan_07}
 Nagao, T. and  Tanaka, T., 
Spectral density of sparse sample covariance matrices, 
{\it J. Phys. A: Math. Theor.},
 {\bf 40} 
 (2007) 
 4973-4987.

\bibitem{newman_06}
 Newman, M. E. J., 
Finding community structure in networks using the eigenvectors of matrices, 
{\it Phys. Rev. E},
 {\bf 74} 
 (2006) 
 036104.

\bibitem{new_gir_02}
 Newman, M. E. J. and  Girvan, M., 
Mixing patterns and community structure in networks, 
{ in: {\it Statistical Mechanics of Complex Networks}, eds. R. Pastor-Satorras, J. Rub\'{\i}\ and A. Diaz-Guilera,}
 (Springer, Berlin, 2003) 
 pp. 66-87.

\bibitem{new_gir_04}
 Newman, M. E. J. and  Girvan, M., 
Finding and evaluating community structure in networks, 
{\it Phys. Rev. E},
 {\bf 69} 
 (2004) 
 026113.

\bibitem{new_bar_wat_06}
 Newman, M.,  Barab\'asi,  A.-L.,  Watts ,  D. J., 
{\it The Structure and Dynamics of Networks}
 (Princeton University Press, Princeton, 2006). 

\bibitem{noh_rie_03}
 Noh, J. C. and  Rieger, H., 
Random walks on complex networks, 
{\it Phys. Rev. Lett.},
 {\bf 92} 
 (2004) 
 118701. 

\bibitem{pal_der_far_vic_05}
 Palla, G.,  Der\'enyi,  I.,  Farkas,  I. and  Vicsek, T., 
Uncovering the overlapping community structure of complex networks in nature and society, 
{\it Nature},
 {\bf 435} 
 (2005) 
 814-818.

\bibitem{ple_gop_ros_am_sta_99}
 Plerou, V.,  Gopikrishnan,  P.,  Rosenow,  B.,  Amaral,  L. A. N. and  Stanley, H. E., 
Universal and nomuniversal properties of cross correlations in financial time series, 
{\it Phys. Rev. Lett.},
 {\bf 83} 
 (1999) 
 1471-1474.

\bibitem{ple_gop_ros_ama_sta_00}
 Plerou, V.,  Gopikrishnan,  P.,  Rosenow,  B.,  Amaral,  L. A. N. and  Stanley, H. E., 
Random matrix approach to cross correlations in financial data, 
{\it Physica A},
 {\bf 287} 
 (2000) 
 374-382.

\bibitem{ple_gop_ros_ama_guh_sta_02}
 Plerou, V.,  Gopikrishnan,  P.,  Rosenow,  B.,  Amaral,  L. A. N.,  Guhr,  T. and  Stanley, H. E., 
Random matrix approach to cross correlations in financial data, 
{\it Phys. Rev. E},
 {\bf 65} 
 (2002) 
 066126.

\bibitem{rad_cas_cec_lor_par_04}
 Radicchi, F.,  Castellano,  C.,  Cecconi,  F.,  Loreto,  V. and  Parisi, D., 
Defining and identifying communities in networks, 
{\it Proc. Natl. Acad. Sci. USA},
 {\bf 101} 
 (2004) 
 2658-2663.

\bibitem{rei_bor_04}
 Reichardt, J. and  Bornholdt, S., 
Detecting Fuzzy Community Structures in Complex Networks with a Potts Model, 
{\it Phys. Rev. Lett.},
 {\bf 93} 
 (2004) 
 218701.

%
%
\bibitem{rheingold_1993}
Rheingold, H.,  {\it The virtual community: Homesteading on the electronic frontier} (Addison-Wesley, Reading, MA, 1993).


\bibitem{rod_bra_88}
 Rodgers, G. J. and  Bray, A. J., 
Density of states of a sparse random matrix, 
{\it Phys. Rev. B},
 {\bf 37} 
 (1988) 
 3557-3562.

\bibitem{rod_aus_kah_kim_05}
 Rodgers, G. J.,  Austin,  K.,  Kahng,  B. and  Kim, D., 
Eigenvalue spectra of complex networks, 
{\it J. Phys. A: Math. Gen.},
 {\bf 38} 
 (2005) 
 9431-9437.

\bibitem{rod_cas_kuh_tak_08}
 Rogers, T.,  Castillo,  I. P.,  K\"uhn,   R. and  Takeda, K., 
Cavity approach to the spectral density of sparse symmetric random matrices, 
{\it Phys. Rev. E},
 {\bf 78} 
 (2008) 
 031116.

\bibitem{saw_amu_sal_ama_09}
 Sawardecker, E. N.,  Amundsen,  C. A.,  Sales-Pardo,  M. and  Amaral, L. A. N., 
Comparison of methods for the detection of node group membership in bipartite networks, 
{\it Eur. Phys. J.},
 {\bf 72} 
 (2009) 
 671-677.

%
%
\bibitem{scott_2000}
Scott, J.,  {\it Social network analysis: A handbook} (Sage, London, Thousand Oaks and New Delhi, 2000).


\bibitem{sem_cug_02}
 Semerjian, G. and  Cugliandolo, L. F., 
Sparse random matrices: the eigenvalue spectrum revisited, 
{\it J. Phys. A: Math. Gen.},
 {\bf 35} 
 (2002) 
 4837-4851.

\bibitem{sim_eri_mas_sne_04}
 Simonsen , I.,  Eriksen ,  K. A.,  Maslov,  S. and  Sneppen, K., 
Diffusion on complex networks: a way to probe their large-scale topological structures, 
{\it Physica A},
 {\bf 336} 
 (2004) 
 163-173.

\bibitem{sla_zha_05}
 Slanina, F. and  Zhang, Y.-C., 
Referee networks and their spectral properties, 
{\it Acta Phys. Pol. B},
 {\bf 36} 
 (2005) 
 2797-2804.

%
%
\bibitem{star_griesemer_1989}
Star, S. L. and Griesemer, J. R.,  Institutional ecology, "translations" and boundary objects: Amateurs and professionals in Berkeley's museum of vertebrate zoology, 1907-39, {\it Social Studies of Science} {\bf 19 }   (1989) 387-420.


\bibitem{strogatz_01}
 Strogatz, S. H., 
Exploring complex networks, 
{\it Nature},
 {\bf 410} 
 (2001) 
 268-276.

\bibitem{tib_one_sar_kas_ker_06}
 Tib\'ely, G.,  Onnela,  J.-P.,  Saram\"aki,  J.,  Kaski,  K. and  Kert\'esz, J., 
Spectrum, intensity and coherence in weighted networks of a financial market, 
{\it Physica A},
 {\bf 370} 
 (2006) 
 145-150.

%
%
\bibitem{wasserman_faust_1994}
Wasserman, S. and Faust, K.,  {\it Social network analysis: Methods and application} (Cambridge University Press Cambridge, 1994).


%
%
\bibitem{wellman_etal_1996}
Wellman, B., Salaff, J., Dimitrova, D., Garton, L., and  AL., E.,  Computer networks as social networks: Collaborative work, telework, and virtual community, {\it Annual Review of Sociology} {\bf 22 }   (1996) 213-238.


%
%
\bibitem{wittel_2001}
Wittel, A.,  Toward a network sociality, {\it Theory, Culture and Society} {\bf 18 }  (2001) 51-76.


%
%
\bibitem{woolgar_2004}
Woolgar, S.,  Reflexive Internet? The British experience of new electronic technologies, in:  {\it The network society: A cross-cultural perspective}, ed. M. Castells (Edward Elgar, Cheltenham and Northampton, MA, 2004) pp. 125-143.


\bibitem{zha_yan_wan_05}
 Zhao, F.,  Yang,  H. and  Wang, B., 
scaling invariance in spectra of complex networks: A diffusion factorial moment approach, 
{\it Phys. Rev. E},
 {\bf 72} 
 (2005) 
 046119. 

\bibitem{zho_ren_med_zha_07}
 Zhou, T.,  Ren,  J.,  Medo,  M. and  Zhang, Y.-C., 
Bipartite network projection and personal recommendation, 
{\it Phys. Rev. E},
 {\bf 76} 
 (2007) 
 046115.

%
%
\end{thebibliography}
\end{document}